\documentclass[a4paper,fleqn,usenatbib]{mnras}

\usepackage{graphicx}	
\usepackage{amsmath}	
\usepackage{amssymb}	

\title[C/O in extrasolar planetesimals]{Carbon to oxygen ratios in extrasolar planetesimals}
\author[Wilson et al.]{David J. Wilson$^1$\thanks{d.j.wilson.1@warwick.ac.uk},
  Boris T. G{\"a}nsicke$^1$, Jay Farihi$^2$, Detlev Koester$^3$ \medskip\\
$^{1}$ Department of Physics, University of Warwick, Coventry CV4 7AL,
UK\\
$^{2}$ University College London, Department of Physics \& Astronomy, Gower Street, London WC1E 6BT, UK \\
$^{3}$ Institut f\"ur Theoretische Physik und Astrophysik, University of Kiel,
24098 Kiel, Germany}

\date{Accepted 2016 April 11. Received 2016 April 11; in original form 2015 August 20}

\pubyear{2016}

\begin{document}
\label{firstpage}
\pagerange{\pageref{firstpage}--\pageref{lastpage}}
\maketitle

\begin{abstract}
Observations of small extrasolar planets with a wide range of densities imply a variety of planetary compositions and structures. Currently, the only technique to measure the bulk composition of extrasolar planetary systems is the analysis of planetary debris accreting onto white dwarfs, analogous to abundance studies of meteorites. We present measurements of the carbon and oxygen abundances in the debris of planetesimals at ten white dwarfs observed with the {\em Hubble Space Telescope}, along with C/O ratios of debris in six systems with previously reported abundances. We find no evidence for carbon-rich planetesimals, with C/O\,$<0.8$ by number in all 16 systems. Our results place an upper limit on the occurrence of carbon-rich systems at $<17$\,percent with a $2\,\sigma$ confidence level.  The range of C/O of the planetesimals is consistent with that found in the Solar System, and appears to follow a bimodal distribution: a group similar to the CI chondrites, with $\log({<}\mathrm{C/O}{>})=-0.92$, and oxygen-rich objects with C/O less than or equal to that of the bulk Earth. The latter group may have a higher mass fraction of water than the Earth, increasing their relative oxygen abundance. 
\end{abstract}


\begin{figure*}
    \centering
    \includegraphics[width=15.0cm]{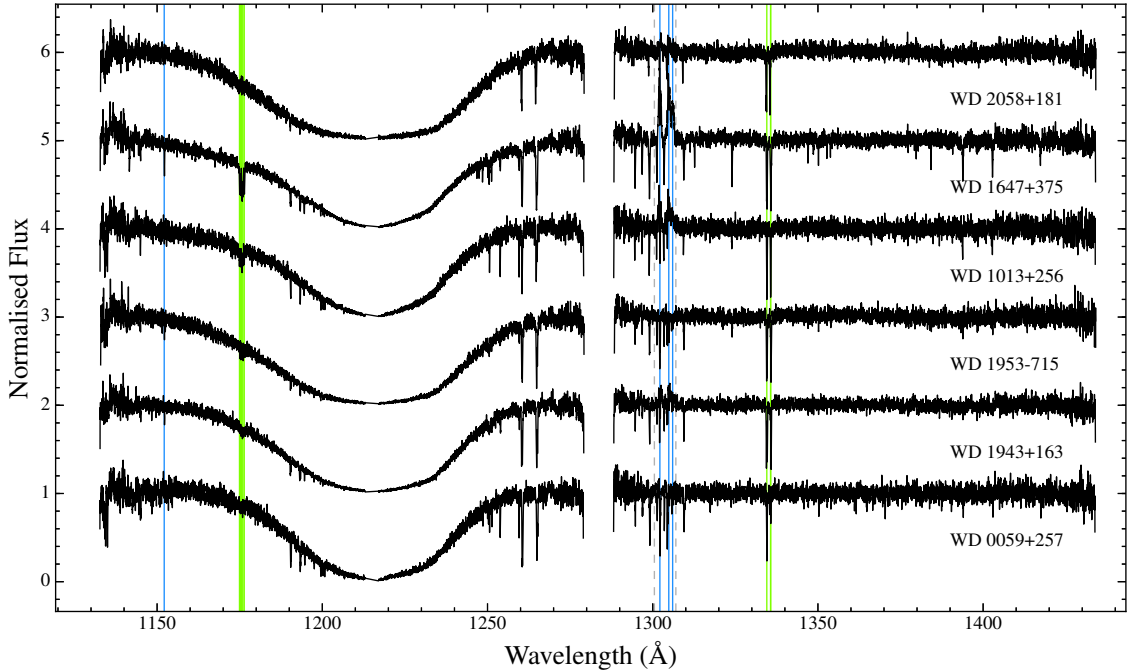}
    \caption{{\em HST}/COS ultraviolet spectra of the first six white dwarfs in Table \ref{tab:new_wds}. Spectra are smoothed with a 5-point boxcar and normalised, then offset by multiples of 1 for clarity. The spectra are dominated by the broad Ly$\alpha$ line (with the central air glow emission line removed), and absorption lines of several metals are present. The wavelengths of the carbon and oxygen absorption transitions are indicated by the green and blue lines, respectively (Table \ref{tab:all_lines}). In some of the spectra, the area between the dashed grey lines is affected by geocoronal oxygen emission, which is not corrected for by the COS pipeline.        \protect\label{fig:spectra}}
\end{figure*}

\section{Introduction}
\label{sec:intro}
The ongoing search for extrasolar planets has been spectacularly successful, with over 1500 confirmed planets discovered to date\footnote{{\tt http://exoplanets.org/}}, including many small objects suspected of being rocky. For a subset of these smallest detected exoplanets, both precision radial velocity measurements and transit photometry have been obtained. This provides a measurement of their masses and radii, and therefore their bulk densities. 
Intriguingly, these densities have a wide spread, and do not follow a simple mass-radius relationship \citep{weiss+marcy14-1,dressingetal15-1}. This may imply that some small exoplanets have compositions distinct from the rocky (and icy) planets and moons of the Solar System, which are all, to first order, a combination of H$_2$O, MgSiO$_3$ and Fe \citep{allegreetal01-1}. Modelling exoplanets with a greater variety of bulk chemistries may account for the differences in bulk densities. However, it is impossible to unambiguously infer the internal composition of  a planet from its density alone. \cite{seageretal07-1} and \cite{sohletal12-1} computed mass-radius relationships for different planetary compositions, finding a significant degeneracy between different densities, interior structures and compositions. 

It has been hypothesized that enhanced C/O levels (relative to the Solar value) in a protoplanetary disc could change the condensation sequence of planetary solids, preferentially forming carbon compounds \citep{kuchner+seager05-1,moriartyetal14-1}. Under conditions where carbon is the most abundant metal, ``carbon planets'' may form. The alternative condensation sequence begins with the formation of CO, incorporating all of the available oxygen and restricting the formation of silicates. Excess carbon then forms SiC and graphite, for example. An Earth-sized carbon planet would likely form with an Fe-rich core, surrounded by a mantle of graphite, carbides and, at higher pressures, diamond. \cite{bondetal10-1} showed that this carbon-based chemistry could become important in protoplanetary discs with C/O\,$\gtrsim0.8$. Carbon could contribute more than half the mass of solid exoplanets formed in such an environment, with only trace oxygen present.

Observational identification of carbon planets is hindered by the inability to measure planetary compositions in-situ, with the exception of the upper atmospheres of a few objects \citep{demingetal13-1,kreidbergetal14-1}. Given the diversity in atmospheric composition between the otherwise chemically similar terrestrial planets in the Solar System, such observations cannot be used to infer the bulk compositions of rocky exoplanets. Neither are the C/O ratios of exoplanet host stars a reliable tracer of disc composition \citep{teskeetal13-1}. Carbon to oxygen ratios in protoplanetary discs computed by \cite{thiabaudeta115-1} show only a weak dependence on the host star abundances. This ratio will also vary within a protoplanetary disc due to regional temperature variations and collisions, amongst other factors \citep{obergetal11-1,gaidos15-1}.

The only method to reliably determine compositions of exoplanetary bodies is via detection of their debris in the photospheres of white dwarfs \citep{zuckermanetal07-1}. Recent studies have shown that 25--50\,percent of all white dwarfs are polluted by debris from planetesimals \citep{zuckermanetal03-1, zuckermanetal10-1,koesteretal14-1,barstowetal14-1}, ranging in mass from small asteroids to objects as large as Pluto \citep{girvenetal12-1,wyattetal14-1}. The bulk composition of these exoplanetary bodies can be inferred from the debris detected in the white dwarf photosphere, analogous to how the compositions of Solar System bodies are inferred from meteorites \citep{lodders+fegley11-1}. High-resolution spectroscopy of over a dozen metal-polluted white dwarfs has revealed accretion of numerous atomic species, allowing detailed studies of the chemical composition of extrasolar planetesimals \citep{kleinetal11-1,gaensickeetal12-1,dufouretal12-1,juraetal12-1,farihietal13-2,xuetal14-1,raddietal15-1, wilsonetal15-1}. Overall, these objects have chemical compositions similar to inner Solar System bodies, dominated by O, Si, Mg and Fe, and volatile depleted \citep{jura+young14-01}. However, the detailed compositions can be very diverse, with objects having enhanced levels of core material \citep{melisetal11-1,gaensickeetal12-1,wilsonetal15-1}, evidence of post-nebula processing \citep{xuetal13-1}, and significant mass fractions of water \citep{farihietal13-2,raddietal15-1}.

Thus far, studies of planetesimal compositions at white dwarfs have predominately focused on individual objects. However, the growing sample of abundance studies now allows conclusions to be derived regarding the overall chemical abundances of (solid) exoplanet precursors in a statistically significant sample of systems. Here, we use these data to constrain the occurrence frequency of carbon planets.

\begin{figure}
    \centering
    \includegraphics[width=8.5cm]{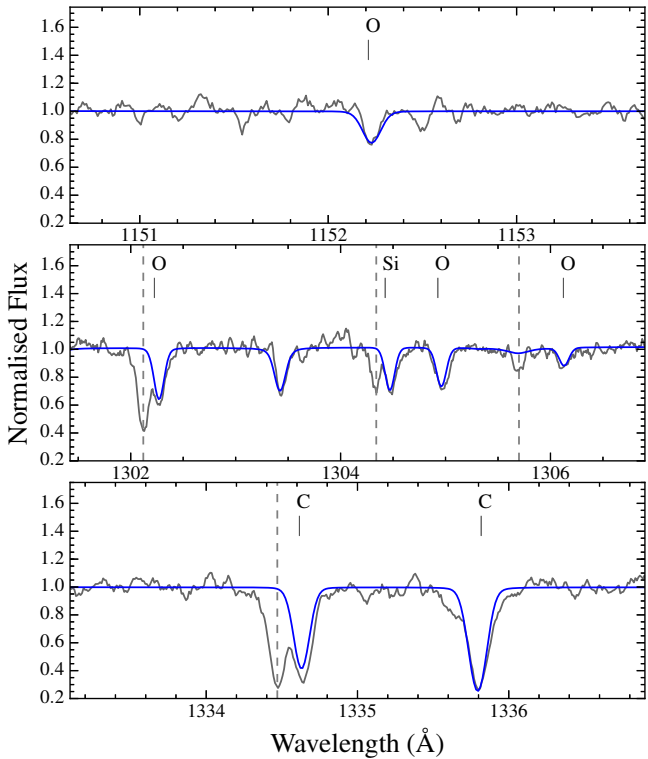}
    \caption{Enlarged sections of the spectrum of WD\,1953--715 showing photospheric \ion{O}{i}\,1152.2, 1302.2, 1304.9, 1306.0\,\AA, \ion{Si}{ii}\,1304.4\,\AA\, and \ion{C}{ii}\,1334.5, 1335.6\,\AA\ absorption lines. The model atmosphere fit used to calculate the abundances is overlaid in blue. Interstellar components of the \ion{O}{i}\,1302.2\,\AA, \ion{Si}{ii}\,1304.4, 1305.6\,\AA\ and \ion{C}{ii}\,1334.5\,\AA\ absorption lines are marked with dashed grey lines.    \protect\label{fig:lines}}
\end{figure}

\begin{table*}
\centering
\caption{New atmospheric parameters and debris accretion rate measurements for ten white dwarfs identified by \citet{koesteretal14-1}. Spectra of the first six are shown in Fig. \ref{fig:spectra}. $^1$Updated from \citet{gaensickeetal12-1} }
\begin{tabular}{lcccc}
\hline\hline
Name & $T_{\mathrm{eff}}~(\mathrm{K})$ & $\log g~(\mathrm{cm\,s^{-2}})$ & $\dot M(\mathrm{C})~(\mathrm{g\,s^{-1}})$ & $\dot M(\mathrm{O})~(\mathrm{g\,s^{-1}})$\\\hline
WD\,2058+181          & $17308\pm235$ & $7.920\pm0.089$ & $(2.06\pm0.95)\times10^6$     & $(1.03\pm0.47)\times10^7$ \\
WD\,1647+375          & $22803\pm310$ & $7.902\pm0.089$ & $(1.14\pm0.52)\times10^7$     & $(2.16\pm0.75)\times10^8$ \\
WD\,1013+256          & $22133\pm301$ & $8.022\pm0.089$ & $(1.92\pm0.66)\times10^6$     & $(3.2\pm1.6)\times10^7$ \\
WD\,1953--715         & $18975\pm258$ & $7.957\pm0.089$ & $(1.56\pm0.72)\times10^6$     & $(2.8\pm1.3)\times10^7$ \\
WD\,1943+163          & $19451\pm264$ & $7.896\pm0.089$ & $(1.40\pm0.64)\times10^6$     & $(1.97\pm0.91)\times10^7$ \\
WD\,0059+257          & $20491\pm278$ & $8.002\pm0.089$ & $\leq2.9\times10^4$  & $(3.4\pm1.6)\times10^7$ \\
PG\,0843+516$^1$      & $22412\pm304$ & $7.902\pm0.089$ & $(2.42\pm1.11)\times10^5$     & $(1.09\pm0.50)\times10^8$ \\
PG\,1015+161$^1$      & $18911\pm257$ & $8.042\pm0.089$ & $\leq6.9\times10^4$  & $(4.9\pm2.3)\times10^7$ \\
SDSS\,J1228+1040$^1$  & $20713\pm281$ & $8.150\pm0.089$ & $(1.70\pm0.78)\times10^5$     & $(4.4\pm2.0)\times10^8$ \\
GALEX\,J1931+0117$^1$ & $21457\pm291$ & $7.900\pm0.089$ & $(7.1\pm4.9)\times10^5$     & $(9.0\pm6.2)\times10^8$ \\
\hline 
\end{tabular}
\label{tab:new_wds}
\end{table*}

\begin{table}
\centering
\caption{List of the absorption lines used for the debris abundance measurements.}
\begin{tabular}{ll}
\hline\hline
Ion & Vacuum rest wavelength (\AA) \\\hline
\ion{C}{ii}  & 1334.530, 1335.660, 1335.708 \\
\ion{C}{iii} & 1174.930, 1175.260, 1175.590, 1175.710, 1175.987, 1176.370 \\
\ion{O}{i}   & 1152.150, 1302.170, 1304.860, 1306.030 \\

\hline 
\end{tabular}
\label{tab:all_lines}
\end{table}

\section{Carbon and Oxygen debris abundances at white dwarfs}
\label{sec:wds}

We present debris abundance measurements for ten white dwarfs observed with the Cosmic Origins Spectrograph on board the {\em Hubble Space Telescope} ({\em HST}/COS) as part of Program IDs 12169, 12869, and 12474 \citep{gaensickeetal12-1,koesteretal14-1}. Table\,\ref{tab:new_wds} presents their effective temperatures  ($T_{\mathrm{eff}}$) and surface gravities ($\log g$), as well as elemental accretion rates. The techniques used to determine these results are described in detail in \cite{koesteretal14-1}, so we only briefly summarise here. Firstly, optical spectra from the SPY survey were refitted with the latest model
grid to determine  temperatures and surface gravities. If no SPY spectra were available, we
used parameters from \cite{gianninasetal11-1}. After correcting for a small systematic difference between the
two determinations, we fixed the surface gravity to the value
obtained from the optical data, and then determined the
temperature from a fit to the ultraviolet COS spectra. For this we used the slope between the optical
photometry and the absolutely calibrated COS spectra as additional
constraint.

The best fit atmospheric parameters were then used to create synthetic spectra containing approximately 14\,000 spectral lines from 14 elements.The atmospheric metal abundances were varied until a good fit was obtained between the synthetic spectra and the observed absorption lines. Adjusting the abundances by $\pm\,0.2$\,dex around the best fit values allowed an estimate of the abundance uncertainties. The uncertainty in the atmospheric parameters has only a small effect on the element abundances (${<}0.04$\,dex). Table \ref{tab:all_lines} lists the absorption lines used  to determine the carbon and oxygen abundances. The oxygen abundances are primarily measured from the \ion{O}{i}\,1152.15\,\AA\ line. The \ion{O}{i} lines around 1300\,\AA\ are affected by geocoronal emission in several of the spectra, which is not corrected for by the COS pipeline. Where no geocoronal emission is present, these lines are still affected by blending with \ion{Si}{ii} and interstellar \ion{O}{i} lines, but still provide (less accurate) abundance determinations which agree with measurements from the \ion{O}{i}\,1152.15\,\AA\ line. 
 
As the metals diffuse out of the white dwarf atmosphere on different time scales, the element abundances in the white dwarf photosphere do not necessarily match those of the debris material. The diffusion time scales were calculated using the same atmospheric models as for the spectral fitting \citep{koester09-1}. As the diffusion time scales for these hydrogen atmosphere white dwarfs are of order days to, at most, months, it is reasonable to assume that the white dwarfs are currently accreting, and accretion and diffusion are in equilibrium. The accretion rate is therefore the ratio of the atmospheric abundance to the diffusion time scale.Radiative levitation, which can change the diffusion time scales or even keep an element in the atmosphere without ongoing accretion \citep{chayer+dupuis10-1}, is taken into account when calculating the diffusion time scales, but has a negligible effect on carbon and no effect on oxygen over the temperature range of our sample.  Finally, the C/O ratio by number is calculated as the ratio of the accretion rates, weighted by the relative atomic masses. 


Analysis of the debris in four of these white dwarfs were presented in \citet{gaensickeetal12-1}, but the abundances used here have been updated with new calculations. Ultraviolet spectra of the remaining six white dwarfs are shown in Fig.\,\ref{fig:spectra}, featuring photospheric absorption lines from a variety of metals, including both carbon and oxygen (Fig.\,\ref{fig:lines}).

In addition to these new measurements, we have assembled all published abundances for carbon and oxygen at white dwarfs both observed with COS and analysed with the same model described above \citep{wilsonetal15-1,xuetal14-1,farihietal13-2, xuetal13-1}. These criteria create a homogeneous sample, which avoids systematic uncertainties that may result from comparing different data sources and models. Where more than one measurement is available we use the most recent result, and we adopt the most commonly used white dwarf designations. In total, we discuss C/O measurements for debris in eleven systems and firm upper limits for another five.

\begin{figure*}
    \centering
    \includegraphics[width=17.0cm]{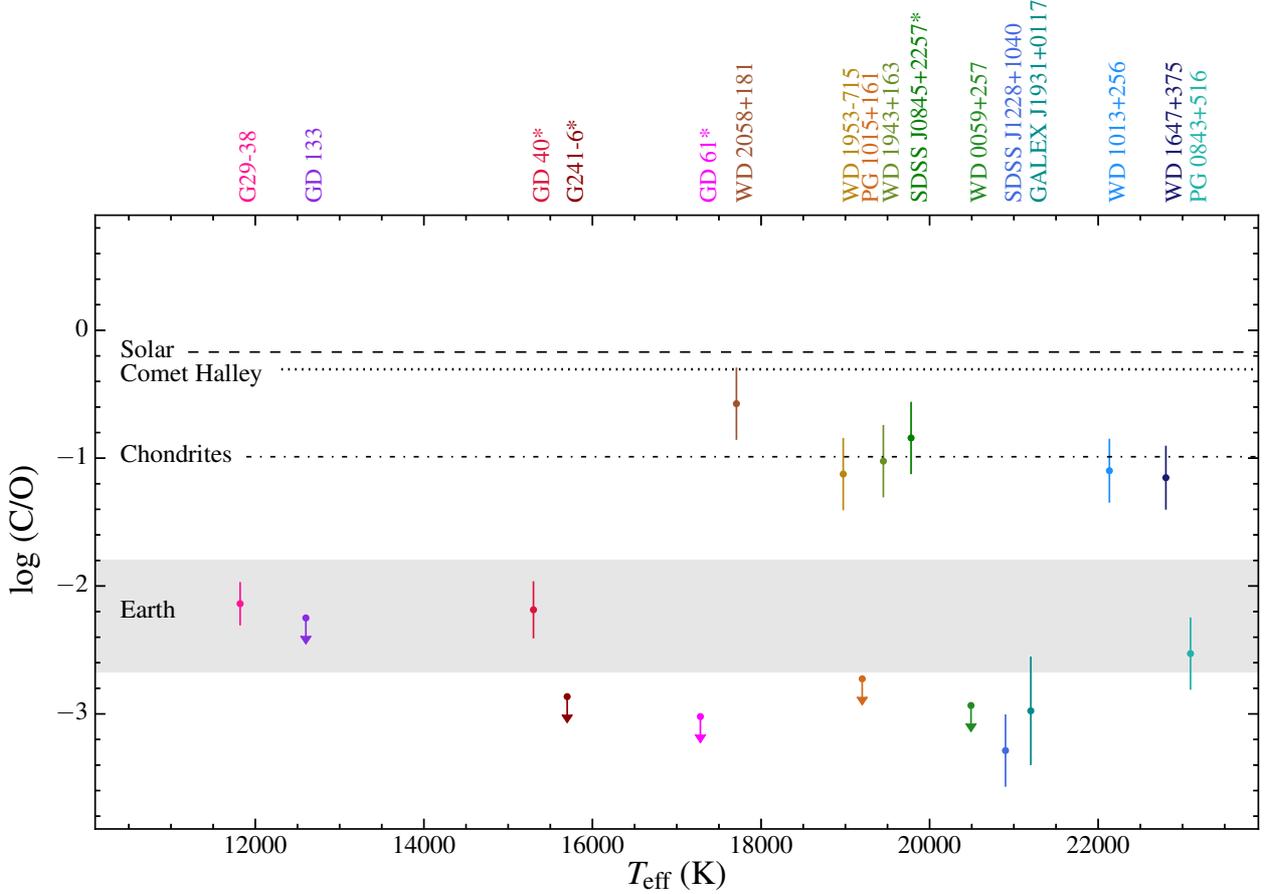}
    \caption{C/O number ratios of the planetesimal debris in our sample, plotted against the effective temperature ($T_{\mathrm{eff}}$) of the host white dwarfs and compared with various Solar System bodies. The colour scheme is intended to aid identification and has no physical significance. White dwarfs with helium atmospheres, which may be affected by convective carbon dredge-up that could enhance their carbon abundances (Sect.\,\ref{sec:wds}), are marked with *. Objects with similar temperatures have been offset slightly for clarity. The shaded area shows the range of values present in the literature for Earth's C/O (Section \,\ref{sec:disc}). \protect\label{fig:c_o}}. 
\end{figure*}

Four of the white dwarfs in our sample have helium dominated atmospheres, labelled in Fig \ref{fig:c_o} and Table \ref{tab:c_o_tab}. These stars develop deep convective envelopes, which may lead to dredge-up of core-carbon into the atmosphere. Dredge-up typically occurs in cool ($T_{\mathrm{eff}}\la12\,000$\,K) white dwarfs, but it has been suggested that a small number of white dwarfs may have helium envelopes thin enough to pollute the atmosphere with core-carbon even at higher temperatures \citep{koesteretal14-2, wilsonetal15-1}. Thus, although we treat the C/O ratios in helium atmosphere white dwarfs as firm detections, this caveat should be kept in mind when discussing the planetary abundances at individual white dwarfs. The majority of our sample (12 out of 16) have hydrogen atmospheres, which are unaffected by dredge up. 

\section{Discussion}
\label{sec:disc}

Figure\,\ref{fig:c_o} and Table\,\ref{tab:c_o_tab} show the C/O ratios of the planetesimal debris at the 16 systems in our sample as a function of effective temperature (and therefore the age since white dwarf formation). We compare these ratios with those for the CI chondritic meteorites \citep{lodders+fegley11-1}, bulk Earth \citep{allegreetal01-1,marty12-1}, Comet Halley \citep{lodders+fegley98-1}, and the Solar photosphere \citep{vonsteiger+zurbuchen16-1}. As carbon chemistry is thought to become an important factor in protoplanetary discs with C/O\,$>0.8$ ($\log\mathrm{(C/O)}>-0.097$), we take this as a lower limit for a planetesimal formed in a carbon-rich environment. We note, however, that planets formed in such discs are predicted to potentially have C/O\,$\gg1.0$ \citep{bondetal10-1}. 

We find no planetary debris with C/O\,$>0.8$. The debris at WD\,2058+181 has the highest ratio, with $\log\mathrm{(C/O)}=-0.57\pm0.28$, still below the Solar value. Applying binomial statistics, we find that planetesimals with C/O\,$>0.8$ occur in $<17$\,percent of systems at a $2\,\sigma$ confidence level, falling to $<6.5$\,percent with $1\,\sigma$ confidence. Our upper limit on high planetary C/O is consistent with that found in stellar abundances by \cite{fortney12-1}, who showed that the fraction of stars with C/O\,$>0.8$ is no more than 10--15\,percent. None of the planetesimal debris in the 16 systems has C/O similar to that of Comet Halley ($\log\mathrm{(C/O)}=-0.04$), supporting the conclusions of \cite{verasetal14-2} that comets are not a significant population of parent bodies for the debris detected at many white dwarfs. There are no observed trends in C/O with the post-main sequence (cooling) age. 

Although none of the systems are carbon-rich, the material does appear to fall into two distinct populations, with an apparent gap between $\log\mathrm{(C/O)}\approx-1$ and $\log\mathrm{(C/O)}\lesssim-2$. Six systems have relatively high C/O ratios, with $\log({<}\mathrm{C/O}{>})=0.12\pm0.07$ (where the error is the $1\sigma$ spread). This is consistent with the CI chondrite meteorites \citep{lodders+fegley11-1}, which are thought to be representative of the primordial composition of the rocky Solar System. It is likely that the debris in these systems originated as small asteroids, which had not undergone significant post-nebula differentiation .

 The remaining ten systems all have C/O less than or equal to that of the bulk Earth. Comparing the relative abundances of carbon and oxygen in this group with the other elements detected in their debris shows that they have a high oxygen abundance (relative to, for example, Si), rather than being relatively poor in carbon. A speculative explanation for this is that the parent bodies of the debris contained a significant amount of water, similar to Ceres or the large moons of the gas giants. High mass fractions of water have already been detected in debris at GD\,61 \citep{farihietal13-2}, which has an upper limit on its C/O ratio placing it in the low C/O group. Addition of water to a planetesimal with an otherwise Earth-like composition would increase the abundance of oxygen, but leave the carbon abundance unchanged, decreasing the C/O ratio. A potential caveat to this argument is the study by \cite{jura+xu12-1} of hydrogen in helium atmosphere white dwarfs in the 80\,pc sample. Their results suggested that water makes up less that one percent of the mass accreting onto the white dwarfs in their sample. However, both the amount and origin of hydrogen in helium atmosphere white dwarfs, and its relevance to debris accretion, are subject to ongoing discussion \citep{koester+kepler15-1, bergeronetal11-1}. 
 
 Additionally, the carbon content of the Earth, and in particular the core, is still subject to discussion. \cite{allegreetal01-1} find a mass fraction of 0.17--0.36\,percent, resulting in a log (C/O) between -1.8 and -2.16. In contrast, \cite{marty12-1} instead calculate a carbon mass fraction of only 0.053\,percent. Using the oxygen fraction from \cite{allegreetal01-1}, this lowers the $\log(\mathrm{C/O}$ to -2.7, consistent with the average of the low C/O systems ($\log({<}\mathrm{C/O}{>})=-2.5\pm0.36$).The range of proposed C/O ratios for Earth is shown by the shaded area in Fig.\ref{fig:c_o}.
 
By providing a strong lower limit of the occurrence of carbon-rich planetesimals, we show that debris-polluted white dwarfs are likely the most powerful diagnostics of carbon chemistry in extrasolar planetesimals, and increasing the sample size will provide stronger constraints on the existence, or lack thereof, of carbon planets. More generally, abundance studies of the debris at white dwarfs are sensitive to a wide variety of elements, making them the ideal tool to systematically investigate the full range of non-gaseous planetary chemistry.

\begin{table}
\centering
\caption{C/O ratios by number shown in Fig.\ref{fig:c_o}, in order of increasing C/O. White dwarfs with helium atmospheres are marked with *. References: 1.\,This work; 2.\,\citet{xuetal13-1}; 3.\,\citet{farihietal13-2}; 4.\,\citet{xuetal14-1}; 5.\,\citet{wilsonetal15-1}. }
\begin{tabular}{lcr}
\hline\hline
Name & $\log\mathrm{(C/O)}$ & Ref.\\\hline 
SDSS\,J1228+1040  & $-3.3\pm0.28$  & 1\\
GD\,61*           & $\leq-3.0$     & 3\\
GALEX\,J1931+0117 & $-3.0\pm0.42$  & 1\\
WD\,0059+257      & $\leq-2.9$     & 1\\
G241-6*           & $\leq-2.9$     & 2\\ 
PG\,1015+161      & $\leq-2.7$     & 1\\
PG\,0843+516      & $-2.5\pm0.28$  & 1\\
GD\,13            & $\leq-2.2$     & 4\\    
GD\,40*           & $-2.2\pm0.22$  & 2\\  
G29-38            & $-2.1\pm0.17$  & 4\\
WD\,1647+375      & $-1.2\pm0.25$  & 1\\
WD\,1013+256      & $-1.1\pm0.25$  & 1\\
WD\,1953-715      & $-1.1\pm0.28$  & 1\\
WD\,1943+163      & $-1.0\pm0.28$  & 1\\
SDSS\,J0845+2257* & $-0.84\pm0.28$ & 5\\
WD\,2058+181      & $-0.57\pm0.28$ & 1\\ 
\hline 
\end{tabular}
\label{tab:c_o_tab}
\end{table}

\section*{Acknowledgements}
The authors thank Yilen Gomez Maqueo Chew  and the anonymous referee for constructive comments, and Mark Hollands for statistics advice.. The research leading to these results has received funding from the European Research Council under the European Union's Seventh Framework Programme (FP/2007-2013) / ERC Grant Agreement n. 320964 (WDTracer). JF gratefully acknowledges the support of the STFC via an Ernest Rutherford fellowship
This paper is based on observations made with the NASA/ESA {\em Hubble Space Telescope}, obtained at the Space Telescope Science Institute, which is operated by the Association of Universities for Research in Astronomy, Inc., under NASA contract NAS 5-26555. These observations are associated with program IDs 12169, 12869 and 12474.




\bibliographystyle{mnras}
\bibliography{aamnem99,aabib} 

\bsp	
\label{lastpage}
\end{document}